\documentstyle[seceq,epsf,wrapft,preprint]{ptptex}

\title{
Coexistence of cluster structure and superdeformation in $^{44}$Ti 
 studied with deformed-basis antisymmetrized molecular dynamics}

\author{
Masaaki {\sc Kimura}$^1$
and Hisashi {\sc Horiuchi}$^{2}$
}

\inst{
$^1$RI-beam Science Laboratory, RIKEN ({\it The Institute of
Physical and Chemical Research}), Wako, Saitama 351-0198, Japan.\\
$^2$Department of Physics, Kyoto University, Kitashirakawa, Kyoto
606-8502, Japan}

\recdate{
}

\abst{
The nucleus $^{44}$Ti has low-lying levels of various kinds of mutually 
very different nuclear structure displaying the richness of the nuclear 
many-body dynamics.  It is shown that the deformed-basis antisymmetrized 
molecular dynamics by the use of the Gogny D1S force reproduces successfully 
and unifiedly two types of coexistence phenomena in $^{44}$Ti. Namely, 
on one hand, the coexistence of the mean field structure and the cluster 
structure is confirmed by verifying the normally deformed structure of 
the $K^\pi$=$3^-_1$ band with a 1-particle-jump intrinsic configuration
and  the $\alpha$+$^{40}$Ca cluster structure of the $K^\pi$=$0^-_2$
band.   The mixed character of the mean-field-like structure and the 
$\alpha$+$^{40}$Ca cluster structure of the ground 
band is also shown.  On the other hand, the coexistence of the normal 
deformed mean field and the superdeformed mean field is confirmed by 
verifying the triaxial superdeformation of the $K^\pi$=$0_2^+$ band and 
the $K^\pi$=$2^+_1$ band which has a 4-particle-jump intrinsic
configuration.   Good reproduction of the experimental data is shown for
many kinds of  quantities including the energy spectra, electric
transition rates,  alpha spectroscopic factors.  Preliminary
discussions are  given on the existence of hyperdeformed excited states,
the relation between  superdeformation and clustering and so on. 
}

\begin{document}

\maketitle

\section{Introduction}
The nucleus $^{44}$Ti is the lightest $N$=$Z$ even-even nucleus in the $pf$ 
shell region and its structure is very important for the study of the 
neutron-proton correlation in the $pf$ shell region. Since we can now 
expect the increasing experimental information about the proton-rich 
nuclei along the $N$$\sim$$Z$ line up to $A$$\sim$100, 
the study of the neutron-proton correlation in 
$^{44}$Ti is even more important than before.  $^{44}$Ti is analogous to 
$^{20}$Ne in the sense that both nuclei have two neutrons and two protons 
outside the doubly closed-shell core nuclei $^{40}$Ca and $^{16}$O, 
respectively.  It is well known\ \cite{supple,kimura} that $^{20}$Ne has 
non-small $\alpha$-clustering character even in the ground band states and 
much prominent $\alpha$+$^{16}$O cluster structure in some of its excited 
bands. Therefore there have been many microscopic theoretical works to 
investigate possible $\alpha$-cluster structure in 
$^{44}$Ti\ \cite{kihara,arima}. However, more convincing 
argument\ \cite{ohkubo} than these microscopic studies about the 
$\alpha$-clustering in $^{44}$Ti was obtained from the knowledge 
of the so-called unique $\alpha$-$^{40}$Ca optical potential which was 
established by the study of the elastic $\alpha$-$^{40}$Ca scattering 
including ALAS (anomalous large angle scattering) and nuclear rainbow 
phenomena\ \cite{delbar}. From the fact that the band head $0^+$ state of 
the lowest rotational band of this unique $\alpha$-$^{40}$Ca optical 
potential is located several MeV below the $\alpha$-$^{40}$Ca threshold, 
the lowest band was assigned to correspond to the ground rotational band 
of $^{44}$Ti\ \cite{ohkubo}, which means that the ground band states of 
$^{44}$Ti should have sizable amount of component of $\alpha$+$^{40}$Ca 
cluster structure.  Furthermore, the $K^\pi$=$0^-$ band which is the 
parity-doublet partner of the lowest band ( $K^\pi = 0^+$ ) was predicted 
to exist with its band head $1^-$ state a few MeV higher above the 
$\alpha$+$^{40}$Ca threshold.  A similar argument was also presented by 
a microscopic approach\ \cite{wada} which fitted $\alpha$-$^{40}$Ca 
scattering data in addition to the usual structure study of the bound 
states of $^{44}$Ti. The parity-doublet partner band with $K^\pi$=$0^-$ 
was actually found later experimentally by using $\alpha$-transfer 
reaction\ \cite{yamaya}. 

It is usually considered that the formation of the mean field becomes 
much more firm in the $pf$ shell and heavier mass region compared with 
the smaller mass region where the $\alpha$ clustering is active especially 
below the mass $A$$\sim$30.  However, recent experiments show that the 
mean field around the mass $A$$\sim$40 is not simply static but dynamic 
in its nature.  It is because recently many low-lying rotational spectra
associated  
with superdefomed bands have been observed in rather light mass regions 
including the mass $A\sim40$ region\ \cite{AR36,CA40,TI44}.  In the case 
of $^{44}$Ti, the rotational band built upon the excited $0^+_2$ state 
at 1.905 MeV\ \cite{simps} was found to extend up to the $J^\pi$=$12^+$ 
state and was discussed to have superdeformed structure\ \cite{TI44}. 
According to the shell model and mean-field-type calculations of 
$^{44}$Ti\ \cite{TI44,quart,zamick,inaku}, this superdeformed 
structure is regarded as having a 4-particle-jump configuration, namely 
8p-4h configuration. The appearance of the superdeformed mean field in 
a very low excitation energy region indicates the presence of the active 
dynamics of the mean-field formation. 

As is shown in the above discussions, $^{44}$Ti is a typical nucleus 
which displays to us vividly the richness and profoundness of the 
nuclear many-body dynamics.  Namely in $^{44}$Ti we have the coexistence of 
the $\alpha$-clustering dynamics and the mean-field dynamics, and also the 
coexistence of the normal deformed mean field and superdeformed mean field.  
Therefore it is an important and challenging task for the nuclear theory 
to describe properly and unifiedly these two kinds of coexistence 
phenomena of the nuclear structure dynamics displayed in $^{44}$Ti.  
We think that the proper description of these two kinds of 
coexistence phenomena in $^{44}$Ti is 
indispensable for us in order to get the proper understanding of the 
neutron-proton correlation in the $pf$-shell region and also in the 
proton-rich region along $N \approx Z$ line.  In our recent 
study of $^{32}$S\ \cite{kimurab,KKH}, we have concluded that the 
superdeformed excited band which is predicted by many mean-field 
calculations\ \cite{S32A,S32B,S32C} to exist with its band-head $0^+$ state 
around 9 MeV above the ground state can be regarded also as being an 
$^{16}$O + $^{16}$O molecular band which is predicted to exist in the same 
excitation energy region from the $^{16}$O-$^{16}$O unique optical potential 
studies\ \cite{ohkubob,kondo}.  This result of ours about $^{32}$S 
demands us that we should study also in $^{44}$Ti the interplay and the 
relation between the cluster structure and the superdeformed mean-field 
structure. 

The purpose of this paper is to study the above-mentioned two kinds of 
coexistence phenomena in $^{44}$Ti by using a single theoretical framework 
of the deformed-basis AMD (antisymmetrized molecular dynamics)
\ \cite{kimura,defamd,KKH}.  The deformed-basis AMD is very suited for 
our present study because the AMD method is a kind of {\it ab initio} 
theory which enables us to study the clustering dynamics without 
assuming the existence of any clusters\ \cite{KKH}, and also because, 
by the use of the deformed basis, the AMD can describe even the 
superdeformed mean field yielding almost the same quantitative results as 
the mean field theory.  We will report that the AMD calculation by the 
use of the Gogny D1S force\ \cite{gogny} surely reproduces and hence 
confirms the existence of the two kinds of coexistence in $^{44}$Ti.  
Namely, firstly as for the confirmation of the coexistence of the 
$\alpha$-clustering dynamics and the mean-field dynamics, we have obtained 
the $K^\pi$=$0^-$ band with prominent $\alpha$+$^{40}$Ca structure and 
the $K^\pi$=$3^-$ band with prominent mean-field-like structure having 
a 1-particle-jump dominant configuration.  Furthermore we have found the 
mixture of the $\alpha$+$^{40}$Ca clustering character and the 
mean-field character in the $K^\pi$=$0^+$ ground band states. 
In the case of the ground state, the calculated $\alpha$+$^{40}{\rm Ca}$
component is about 40\%. When we judge from the usual expectation that
in $^{44}$Ti the mean field 
is formed more firmly than in $^{20}$Ne and also the spatial-symmetry-
breaking (hence  $\alpha$-cluster-breaking) spin-orbit force has stronger 
effects than in $^{20}$Ne, it seems that we may doubt the $\alpha$-
clustering character in the ground band.  However, the above-mentioned 
result of our calculation gives us an important theoretical confirmation 
of the non-small $\alpha$+$^{40}$Ca clustering character of the ground 
band, which is consistent with the study by the use of the unique 
$\alpha$-$^{40}$Ca optical potential mentioned before.   Just like we have 
obtained the $K^\pi$=$3^-$ band which is formed by the excitation of the 
single-particle degree of freedom from the mean-field-structure component 
contained in the ground band, we have obtained, on the other hand, an 
excited band which is formed by the excitation of the inter-cluster 
relative motion from the $\alpha$+$^{40}$Ca-structure component 
contained in the ground band.  This excited band with $K^\pi$=$0^+$ 
has been already reported to be observed experimentally and our AMD 
results are consistent to this experimental report. 
Secondly, as for the confirmation of the coexistence of the normal 
deformed mean field and superdeformed mean field, 
we have obtained, in addition to the normal deformation of the ground 
band and the $K^\pi$=$3^-$ band, the superdeformation of the 
$K^\pi$=$0^+_2$ band upon the $0^+_2$ state which has 4-particle-jump 
intrinsic configuration with large deformation parameter $\beta$$\sim$0.5
and with large $\gamma$ parameter $\gamma$$\sim$$25^\circ$.  Due to this 
triaxiality of the superdeformation we have also obtained a
$K^\pi$=$2^+_1$ superdeformed side band which we consider to correspond
to the observed $K^\pi$=$2^+_1$ band upon the $2^+_3$ state at 2.88 MeV.   
The triaxiality of the $^{44}$Ti superdeformation was also 
reported in Ref.\ \cite{inaku}. 

The contents of this paper are as follows: In the next section (section 
2), we briefly explain the formalism of the deformed-basis AMD 
and the generator coordinate calculation in which the constrained 
deformed-basis AMD wave functions are superposed after the angular 
momentum projection. In section 3, we discuss the energy surfaces with 
good spins as functions of the deformation parameter $\beta$ and 
the comparison of calculated energy spectra with experiments.  The 
density distributions of the obtained intrinsic wave functions are
discussed.  The existence of the highly excited states with
hyperdeformed structure  is also suggested. 
In section 4, we discuss the obtained level scheme of $^{44}{\rm Ti}$
which includes many kinds of the nuclear structure. Firstly, we explain
the existence of the different kinds of the nuclear excitations in this
nucleus together with making the comparison with the available
experimental data such as the excitation energies and the $E2$
transition probabilities. Then we focus on the $\alpha$+$^{40}{\rm Ca}$
clustering in the obtained level scheme. The $\alpha$+$^{40}{\rm Ca}$
component in the ground band is analyzed and the excitations of the
inter-cluster motion in the excited bands are discussed. 
Section 5 is for summarizing discussions.

\section{Framework of the deformed-basis antisymmetrized molecular
 dynamics plus generator coordinate method}
In this section, the framework of the deformed-basis AMD+GCM is explained
briefly. For more detailed explanation of the framework of the
deformed-basis AMD, readers are referred to references
\cite{KKH,kimura}. 
The intrinsic wave function of the system with
mass A is given by a Slater determinant of single-particle wave packets
$\varphi_i({\bf r})$; 
\begin{eqnarray}
 \Phi_{int} &=& \frac{1}{\sqrt{A!}}\det
  \{\varphi_1,\varphi_2,...,\varphi_A \} ,\label{EQ_INTRINSIC_WF}\\
 \varphi_i({\bf r}) &=& \phi_i({\bf r})\chi_i\xi_i ,
\end{eqnarray}
where the single-particle wave packet $\varphi_i$ consists of the
spatial $\phi_i$, spin $\chi_i$ and isospin $\xi_i$
parts. Deformed-basis AMD employs the triaxially deformed Gaussian
centered at ${\bf Z}_i$ as the spatial part of the single-particle wave
packet. 
\begin{eqnarray}
 \phi_i({\bf r}) &\propto& \exp\biggl\{-\sum_{\sigma=x,y,z}\nu_\sigma
  (r_\sigma - {\rm Z}_{i\sigma})^2\biggr\},\nonumber\\
 \chi_i &=& \alpha_i\chi_\uparrow + \beta_i\chi_\downarrow,
  \quad |\alpha_i|^2 + |\beta_i|^2 = 1\nonumber \\
 \xi_i &=& proton \quad {\rm or} \quad neutron. \label{EQ_SINGLE_WF}
\end{eqnarray}
Here, the complex number parameter ${\bf Z}_i$ which represents the centroid
of the Gaussian in the phase space takes independent value for each
nucleon. The width parameters $\nu_x, \nu_y$ and $\nu_z$ are real number
parameters and take independent values for $x$, $y$ and $z$ directions,
but are common to all nucleons. Spin part $\chi_i$ is parametrized by
$\alpha_i$ and $\beta_i$ and isospin part $\xi_i$ is fixed to up
(proton) or down (neutron). ${\bf Z}_i$, $\nu_x, \nu_y, \nu_z$ and
$\alpha_i$, $\beta_i$ are the variational parameters and optimized by
the method of frictional cooling. The advantage of the
triaxially deformable single-particle wave packet is that it makes
possible to describe the cluster-like structure and deformed mean-field
structure within a single framework which is discussed in detail in
reference\cite{kimura}.  

As the variational wave function, we employ the parity projected wave
function as in the same way as many other AMD studies
\begin{eqnarray}
 \Phi^{\pm} = P^\pm \Phi_{int} = \frac{(1\pm P_x)}{2} \Phi_{int} ,\label{EQ_PARITY_WF}
\end{eqnarray}
here $P_x$ is the parity operator and $\Phi_{int}$ is the intrinsic wave
function given in Eq(\ref{EQ_INTRINSIC_WF}). Parity projection  makes
possible to determine the different structure of the intrinsic state for
the different parity states. 

Hamiltonian used in this study is as follows;
\begin{eqnarray}
\hat{H} = \hat{T} + \hat{V_n} + \hat{V_c} - \hat{T_g} ,
\end{eqnarray}
where $\hat{T}$ and $\hat{T}_g$ are the total kinetic energy and the
energy of the 
center of mass motion, respectively.
We have used the Gogny force with D1S parameter set as an effective
nuclear force $\hat{V}_n$. Coulomb force $\hat{V}_c$ is approximated by
the sum of seven Gaussians. 
The energy variation is made under the constraint on the nuclear
quadrupole deformation by adding to $\hat{H}$ a constraint potential
$V_{cnst}=v_{cnst}(\langle\beta\rangle^2-\beta_0^2)^2$ with a large
positive value for 
$v_{cnst}$. At the end of the variational calculation, the expectation
value of $V_{cnst}$ should be zero in principle and in the practical
calculation, we confirm it is less than 0.1keV. The optimized wave
function is denoted by $\Phi^{\pm}(\beta_0)$. Here it should be noted
that this constraint does not refer to the deformation parameter
$\gamma$, which means that $\Phi^{\pm}(\beta_0)$ with positive $\beta_0$
can be not only prolate but also oblate. 

From the optimized wave function, we project out the eigenstate of the
total angular momentum $J$,
\begin{eqnarray}
 \Phi^{J\pm}_{MK}(\beta_0) = P^{J}_{MK}\Phi^{\pm}(\beta_0)
  = P^{J}_{MK}P^{\pm}\Phi_{int}(\beta_0).
  \label{EQ_ANGULAR_WF}
\end{eqnarray} 
Here $P^{J}_{MK}$ is the total angular momentum projector. The integrals
over the three Euler angles included in the $P^{J}_{MK}$ are evaluated by
the numerical integration. 

Furthermore, we superpose the wave functions $\Phi^{J\pm}_{MK}$ which
have the same parity and the angular momentum but have different value of
deformation parameter $\beta_0$ and $K$. Thus the final wave function of
the system becomes as follows;
\begin{eqnarray}
 \Phi_n^{J\pm} = c_n\Phi^{J\pm}_{MK}(\beta_0)
  + c_n^\prime\Phi^{J\pm}_{MK^\prime}(\beta_0^\prime) + \cdots,
  \label{EQ_GCM_WF}
\end{eqnarray}
where other quantum numbers except total angular momentum and parity
are represented by $n$. The coefficients $c_n$, $c'_n$,... are determined
by the Hill-Wheeler equation,
\begin{eqnarray}
 \delta \bigl(\langle\Phi^{J\pm}_n|\hat{H}|\Phi^{J\pm}_n\rangle - 
  E_n \langle\Phi^{J\pm}_n|\Phi^{J\pm}_n\rangle \bigr) =0.
  \label{EQ_GCM_EQ}
\end{eqnarray}

In order to analyze the obtained wave function
$\Phi^{J^\pm}_{MK}(\beta_0)$, we extract the single particle structure of
$\Phi_{int}(\beta_0)$ by constructing the Hartree-Fock Hamiltonian from
$\Phi_{int}(\beta_0)$ as follows\cite{dote}. 
When the optimized wave function 
$\Phi^{\pm}_{int}(\beta_0)=P^{\pm}\frac{1}{\sqrt{A!}}\det \{\varphi_1,\varphi_2,...,\varphi_A
\}$  is given, we calculate the orthonormalized basis $\phi_\alpha$
which is a linear combination of the single-particle wave packets
$\varphi_i$, 
\begin{eqnarray}
 \phi_\alpha=\frac{1}{\sqrt{\mu_\alpha}}\sum_{i=1}^A c_{i\alpha} \varphi_i.
\end{eqnarray}
Here, $\mu_\alpha$ and $c_{i\alpha}$ are the eigenvalue and eigenvector
of the overlap matrix $B_{ij}\equiv\langle\varphi_i|\varphi_j\rangle$,
\begin{eqnarray}
 \sum_{j=1}^{A}B_{ij}c_{j\alpha}=\mu_\alpha c_{i\alpha},
\end{eqnarray}
and it is clear that $\{\phi_\alpha\}$ are orthonormalized from this
relation. Using this basis set of $\phi_\alpha$, we calculate the
Hartree-Fock single-particle Hamiltonian $h_{\alpha\beta}$ which is
defined as, 
\begin{eqnarray}
 h_{\alpha\beta} &\equiv& \langle \phi_\alpha|\hat{t}|\phi_\beta\rangle 
+ \sum_{\gamma=1}^A 
\langle\phi_\alpha\phi_{\gamma}|\hat{v}|
\widetilde{\phi_\beta\phi_\gamma}\rangle \nonumber\\
&&+\frac{1}{2}\sum_{\gamma\delta}\langle\phi_{\gamma}\phi_{\delta}|\phi^*_\alpha\phi_\beta\frac{\partial \hat{v}}{\partial \rho}|\widetilde{\phi_\gamma\phi_\delta}\rangle. \label{eq::hf_hamiltonian}
\end{eqnarray}
By the diagonalization of $h_{\alpha\beta}$, we obtain the
single-particle energy $\epsilon_s$ and single-particle wave function
$\tilde{\phi}_s$. 
\begin{eqnarray}
  \sum_{\beta=1}^{A}h_{\alpha\beta} f_{\beta s} =  \epsilon_s f_{\alpha s},\\
 \tilde{\phi}_s = \sum_{\alpha=1}^{A} f_{\alpha s}\phi_\alpha. 
\end{eqnarray}
We note that the single-particle energy $\epsilon_s$ and wave function
$\widetilde{\phi}_s$ are obtained 
for occupied states but not for unoccupied states from this
method. Furthermore, since the actual variational calculation is made
after the parity projection (the superposition of the two
Slater determinants), it does not allow the naive interpretation of
$\phi^\pm(\beta_0)$ by the single-particle picture. However, the
single-particle structure obtained by this method is useful to
estimate the particle-hole structure of the obtained wave
function. From the parity $\pi_s^\pm=\langle\widetilde{\phi}_s|P^\pm
|\widetilde{\phi}_x\rangle$ and the the angular momentum
$\langle\widetilde{\phi}_s|\hat{l}^2|\widetilde{\phi}_s\rangle$, we have
estimated the particle-hole structure which is denoted as
`$n\hbar\omega$ structure' in the following. 

\section{Energy curves of the positive- and negative- parity states}
\subsection{Positive parity states}
\begin{figure}
\epsfxsize =\hsize
\epsfbox{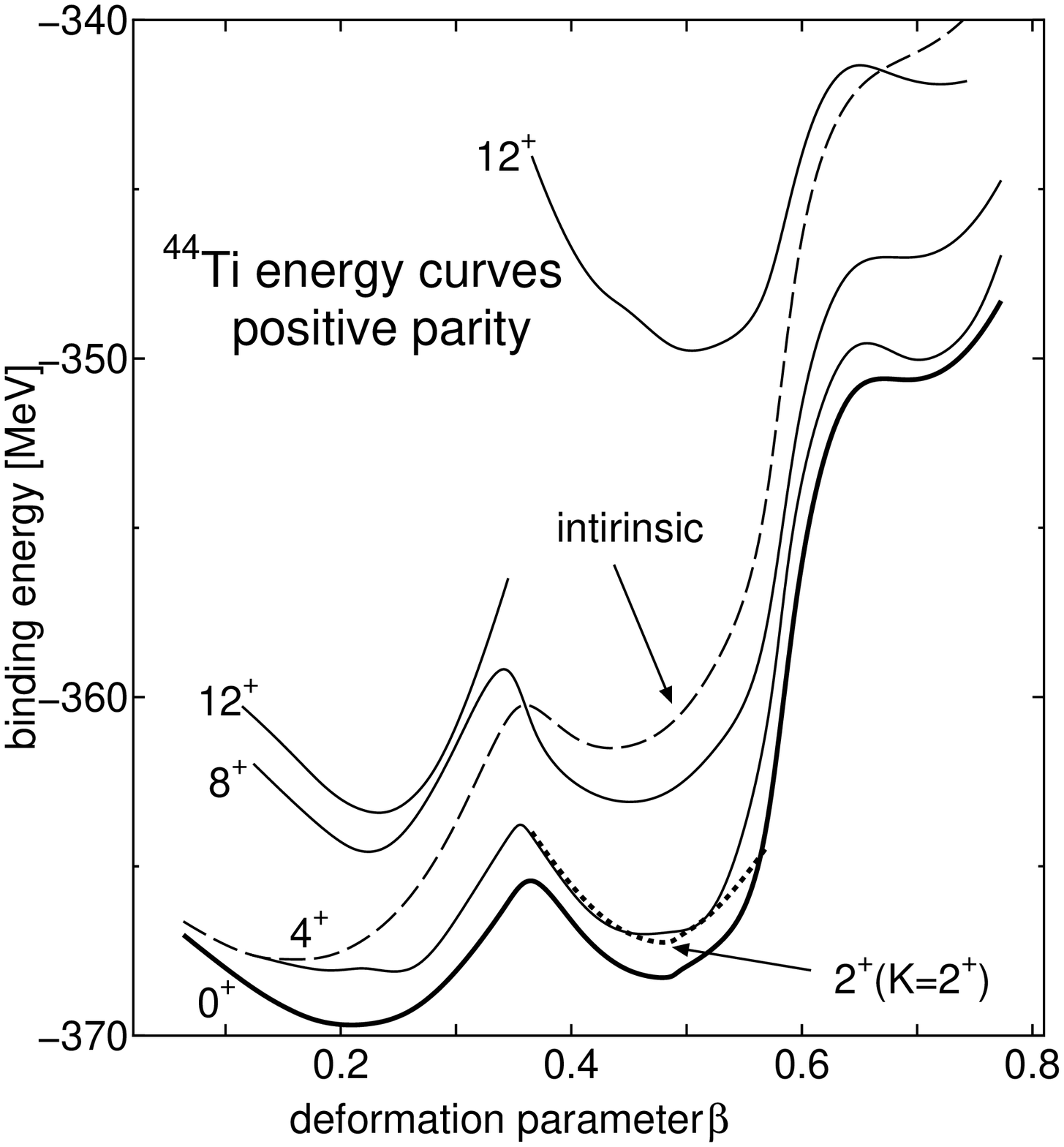}
\caption{Energy curves for each spin state as functions of the
 quadrupole deformation  parameter $\beta$. $0^+$, $4^+$, $8^+$ and
 $12^+$ states are shown for the presentation. The $2^+$ state which is
 the  band head state of $K^\pi$=$2^+$ band is shown by the dotted
 line. The parity-projected intrinsic  state (before the angular
 momentum projection) is also shown by the dashed  line. } 
\label{fig::positive_surface}
\end{figure}

\begin{figure}
\epsfxsize =\hsize
\epsfbox{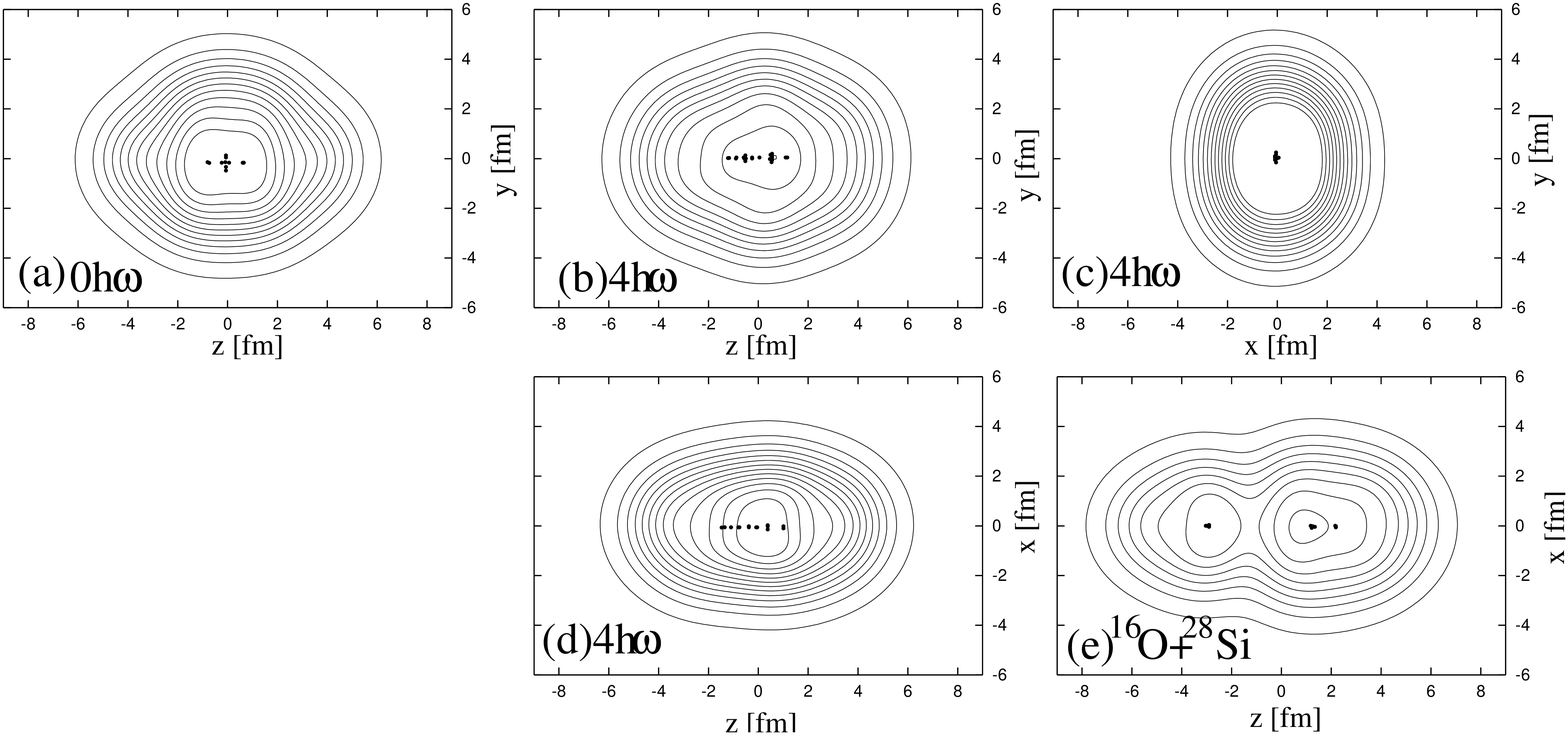}
\caption{Density distributions of the intrinsic wave functions obtained
 by the variational calculation after the projection to positive
 parity. (a) shows the density distribution for the $0\hbar\omega$
 configuration. (b), (c) and (d) show that for the $4\hbar\omega$
 configuration plotted from different angles. (e) shows that for the
 $^{16}{\rm O}$+$^{28}{\rm Si}$ configuration. Black points in these
 figures show the forty-four centroids of the single particle wave
 packets  ($\Re({\bf Z}_i)$) in the coordinate space.}
\label{fig::positive_density}
\end{figure}

After the variation and angular momentum projection, we obtain the
energy curves as functions of the matter quadrupole deformation
parameter $\beta$. Before discussing the low-lying states of $^{44}{\rm
Ti}$, we study the features of the energy curves to clarify the
characters of the wave functions on the energy curves. The
positive-parity energy curve (figure \ref{fig::positive_surface}) has three
energy minima in each spin state and 
the wave functions at these minima have different particle-hole
configurations. The wave functions around the minimum at $\beta$=0.2
have the $0\hbar\omega$ configuration and they 
contribute to the ground band. Each spin state has a parity symmetric
intrinsic density distribution and does not show the apparent existence
of an $\alpha$+$^{40}{\rm Ca}$ cluster correlation. The centroids of
the single particle wave packets are gathered around the center of the
nucleus, that is in contrast to the case of the $^{20}{\rm Ne}$ ground
band in which single particle wave packets are separated into two parts
(4+16) describing the spatially separated $\alpha$ and $^{16}{\rm O}$
subunits. But as we will discuss in the next section, each spin states
(except $10^+$ and $12^+$) around $\beta$=0.20 contains
$\alpha$+$^{40}{\rm Ca}$  
cluster correlation. Around $\beta$=0.35, the change of the single
particle configuration occurs and the wave function  has a $4\hbar\omega$
configuration in which four nucleons are excited into the $pf$-shell
from $^{40}{\rm Ca}$ core. This type of the configuration has its energy
minimum around $\beta$=0.5. Interestingly, the excitation energy of this
$4\hbar\omega$ state with respect to the $0\hbar\omega$ state is quite
small after the 
angular momentum projection. In the present result, it is only
1.7MeV in the case of the $0^+$ state. The angular momentum projection
lowers the energy of this state by about 6MeV. This state corresponds to
the band head state of the superdeformed $K^\pi$=$0^+_2$ band which
starts from the $0^+_2$ state at 1.9 MeV in experiments. Furthermore,
this state has a triaxially deformed 
intrinsic density distribution ($\gamma$=0.25) as is shown in
figure \ref{fig::positive_density}. The triaxial deformation of this
state have been also pointed out by the Hartree-Fock
calculation\cite{inaku} which is also free from any assumptions on the
deformation symmetry of wave function. When  
we perform angular momentum projection, this triaxial deformation
produces the $K^\pi$=$2^+$ band together with the $K^\pi$=$0^+_2$ band
mentioned above. In figure \ref{fig::positive_surface}, only the $2^+$
state is plotted among the $K^\pi$=$2^+$ band members to avoid the
unsightliness of this figure. 
At larger deformation, the energy of the system
increases rapidly, but again it has third local minimum around
$\beta$=0.7.  Its energy is higher by about 19MeV than the
$0\hbar\omega$ state in the case of $0^+$ state. Though the quadrupole
deformation is not large enough, this   state can be regarded as the
hyperdeformed state, if we regarded the $4\hbar\omega$ state around
$\beta$=0.5 as the superdeformed state. Because of the large parity
asymmetry of the wave function, the system has the parity mixed single
particle orbits and it makes hard to define the particle-hole structure
of the usual single particle picture. It is notable that the density
distribution of this state shows the possible relationship with the
$^{16}{\rm O}$+$^{28}{\rm Si}$ molecular structure. 
Different from other states, the centroids of the single particle wave
packets are separated into two parts (16 packets at left side and 28
packets at right side in figure \ref{fig::positive_density} (e))
describing spatially separated $^{16}{\rm O}$ and 
$^{28}{\rm Si}$. 
\begin{figure}
\epsfxsize =\hsize
\epsfbox{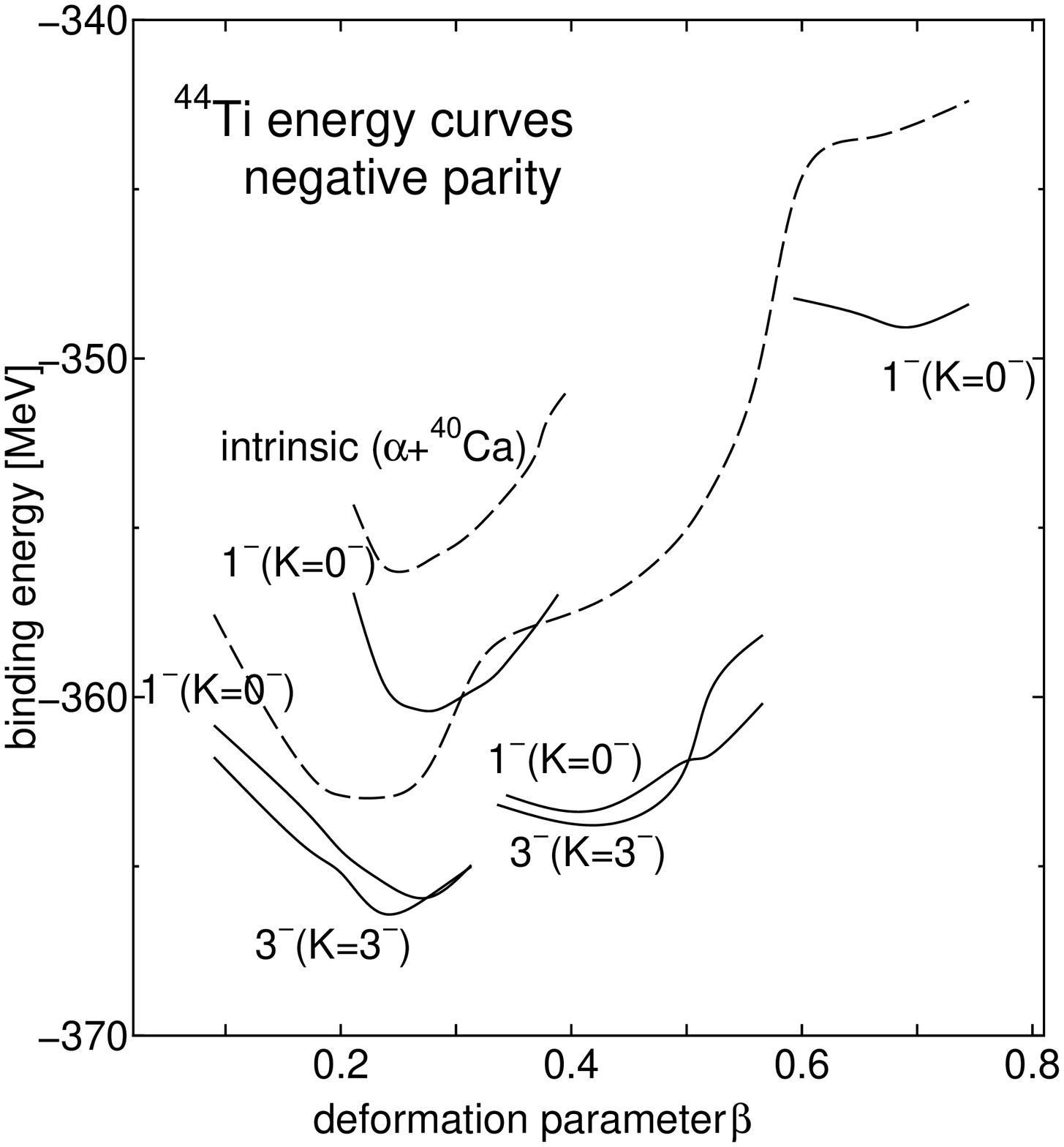}
\caption{Energy curves for each spin state as functions of the
 quadrupole deformation parameter $\beta$. Only the band head states are
 shown for the presentation. The parity-projected intrinsic 
 state (before the angular momentum) is also shown by the dashed line.}
\label{fig::negative_surface}
\end{figure}
\subsection{Negative parity states}
Next, we discuss the negative-parity energy curves (figure
\ref{fig::negative_surface}) whose behavior is 
rather complicated than that of the positive-parity states.  In the
negative-parity states of this nucleus, the intrinsic wave functions have
parity mixed single particle orbits. Therefore, we cannot conclude
the definite $\hbar\omega$ structure, though for the presentation we
denote these structure with the $\hbar\omega$ notation which is roughly
estimated from the parity of the single particle orbit. In the small
deformation region, we obtained the 
$K^\pi$=$3^-$ and $K^\pi$=$0^-$ bands and both have minima around
$\beta$=0.25.  Roughly speaking, the intrinsic wave functions of these
states have $1\hbar\omega$ structure. The $K^\pi$=$3^-$ band corresponds
to the observed lowest negative-parity $K^\pi$=$3^-$ band, while the existence
of the $K^\pi$=$0^-$ band in this energy region has not been
established in experiments.  In the medium deformed region around
$\beta$=0.4, we have 
obtained other $K^\pi$=$0^-$ and $K^\pi$=$3^-$ bands which are also not
established in  experiments. In these bands, $3\hbar\omega$ and
$5\hbar\omega$ 
structures are mixed and they can be regarded as the $1\hbar\omega$
de-excited and excited states of the triaxial superdeformed state with
the positive-parity, although their triaxiality are smaller than that of
the positive-parity states. Their triaxial deformation are about
$\gamma$=$0.16^{\circ}$ at most.   
\begin{figure}
\epsfxsize =\hsize
\epsfbox{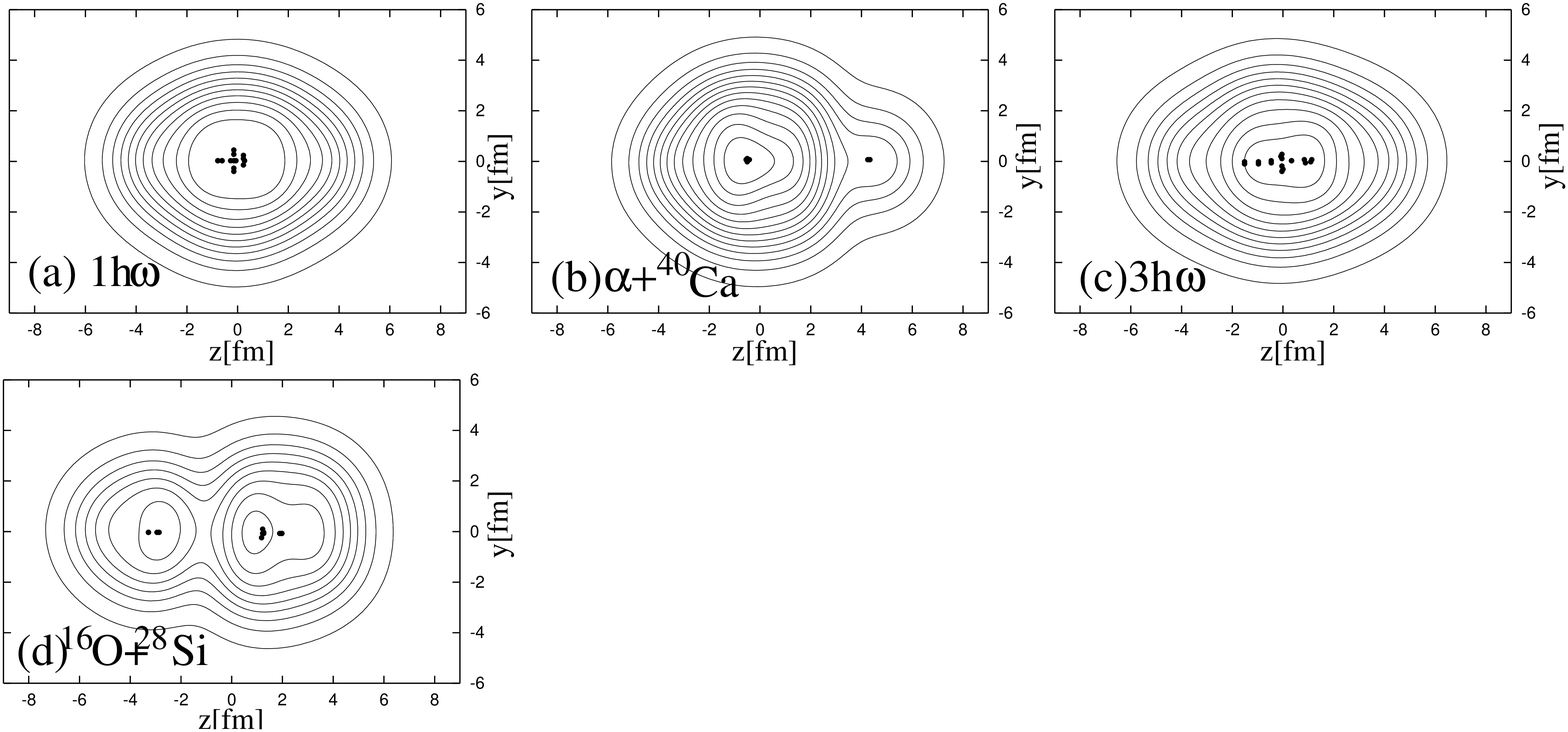}
\caption{Density distributions of the intrinsic wave functions obtained
 by the variational calculation after the projection to negative
 parity. (a) and (c) shows the density distribution for the $1\hbar\omega$
 and $3\hbar\omega$ configurations, respectively. (b) shows that for the
 $\alpha$+$^{40}{\rm Ca}$  configuration.  
 (d) shows that for the $^{16}{\rm O}$+$^{28}{\rm Si}$
 configuration. Black points in these  figures show the forty-four
 centroids of the single particle wave  packets  ($\Re({\bf Z}_i)$) in
 the coordinate space. }  
\label{fig::negative_density}
\end{figure}

Interestingly, we have also found the local minimum between the
$1\hbar\omega$ structure and $3\hbar\omega$ ($5\hbar\omega$) structure 
which is located around $\beta$=0.27. This intrinsic state is 
obtained by performing the frictional cooling calculation using the
$\alpha$+$^{40}{\rm Ca}$ Brink wave function as the initial wave
function. In this calculation, this 
structure continues to be stable during the cooling variational
calculation after the 
parity projection to the negative-parity, although of course, the wave
function considerably deviates from the initial Brink wave function
through the cooling. It is contrasting to the case of the
positive-parity state, in which the $\alpha$ particle is absorbed into
the $^{40}{\rm Ca}$ core even when we start the cooling from the
$\alpha$+$^{40}{\rm Ca}$ Brink wave function. This state corresponds to
the theoretically suggested\cite{ohkubo,wada} and experimentally
observed\cite{yamaya} $\alpha$+$^{40}{\rm Ca}$ cluster structure with 
the negative-parity. However, we will see in the next section that this
band is strongly mixed with the above mentioned $K^\pi$=$0^-$ band
$(3\hbar\omega$ and $5\hbar\omega)$ after the GCM calculation and this
mixing causes the 
fragmentation of the $\alpha$+$^{40}{\rm Ca}$ rotational band states
with negative-parity. In the largely deformed region, we found the
counter part which originates from the parity asymmetry of the
$^{16}{\rm O}$+$^{28}{\rm Si}$ structure obtained in the positive-parity 
state. The $^{16}{\rm O}$+$^{28}{\rm Si}$ structure with the negative
parity gives rise to a $K^\pi$=$0^-$ band.     

%

\section{Variety of the nuclear structure in the low-lying states of
 $^{44}{\rm Ti}$}
\begin{figure}
\epsfxsize =\hsize
\epsfbox{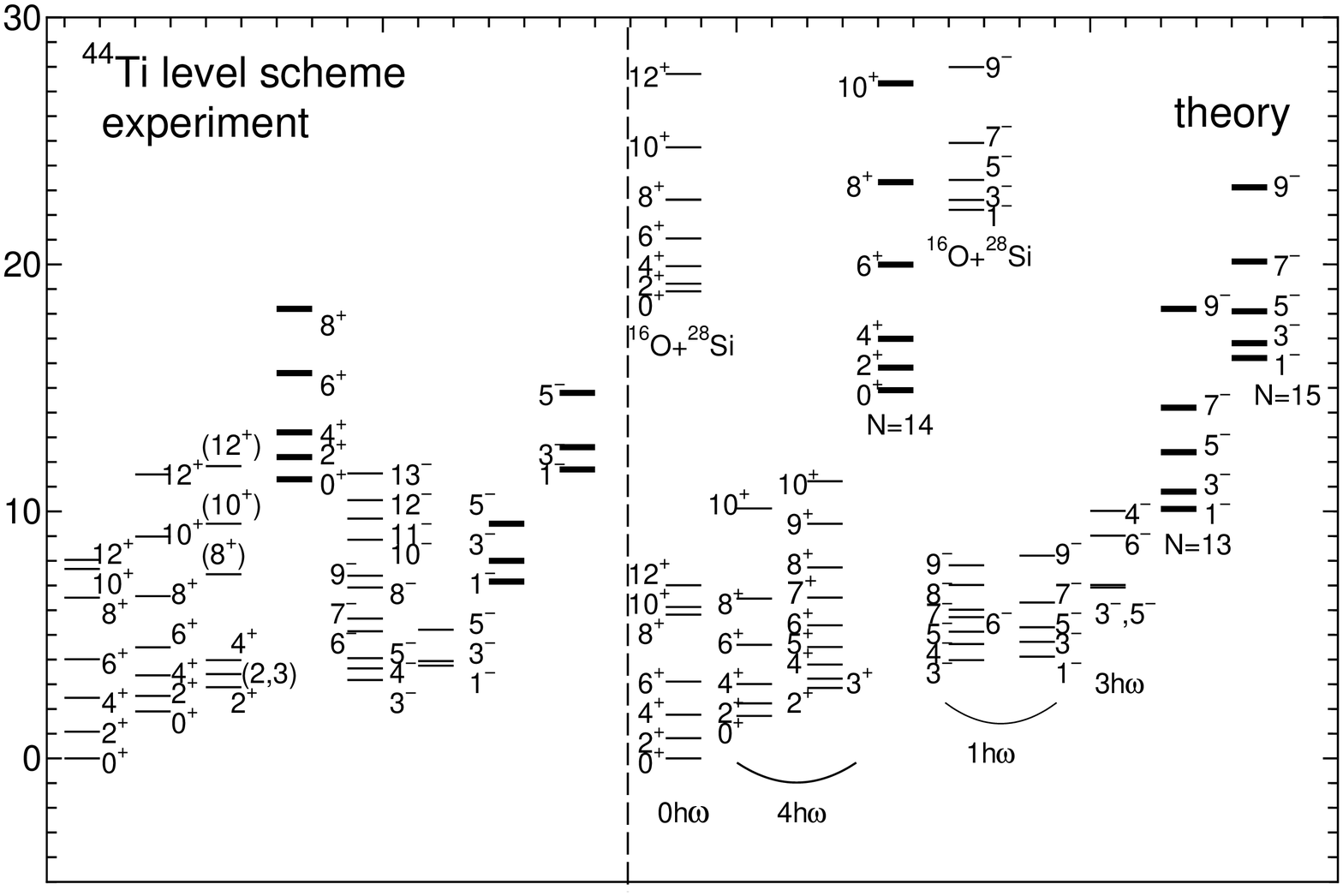}
\caption{Observed and calculated level scheme of $^{44}{\rm Ti}$. 
Only the observed states which correspond to the  calculated states are
shown. In the calculated results, the main component of each band is
shown below the band head state (ex. $0\hbar\omega$) as a guide. The
 states plotted by the bold lines in both experiment 
and theory correspond to the $\alpha$+$^{40}{\rm Ca}$ cluster
states. These states except of N=14 band in the theoretical result are
shown by the averaged energy of the fragmented states (see next
subsection).} 
\label{fig::level}
\end{figure}

The calculated and observed level schemes of $^{44}{\rm Ti}$ are
presented in figure \ref{fig::level}. In this section, we discuss the
obtained level schemes together with other calculated quantities such as
the $E2$ transition probabilities. Firstly, we discuss the different
characteristics of the ground and excited states. Then we focus on the
$\alpha$+$^{40}{\rm Ca}$ clustering in this nucleus. We will see that
the observed eight rotational bands shown in figure \ref{fig::level} are
all reproduced well by the theory.
\subsection{Low-lying level scheme}
\subsubsection{$K^\pi$=$0^+_1$ ground band}
The ground band member states mainly consist of the $0\hbar\omega$
structure and are obtained up to the $12^+$ state. The obtained binding
energy of the ground state is -372.4MeV which underestimates the
observed value by 3 MeV. The excitation
energies of the ground band member states always  
underestimate the experimental values slightly. One of the reason of
this deviation may be the pairing correlation which may not be
included sufficiently in the present calculation, because it will act most
strongly in the ground state. Of course, the insufficiency
of the deformed-basis AMD wave function and/or the insufficiency of the
Gogny D1S force to describe this band would be other reasons, since the
excitation energy of the $12^+$ state is too low to be explained by only 
the underestimation of the pairing correlation. Nevertheless we
can appreciate the description of 
the ground band by the present model. The intra-band $E2$ transition
probabilities of the ground band are reproduced well without any
effective charge (table\ref{tab::ground_E2}). It is to be noted that the
calculated excitation energies show a similar characteristic behavior
to the experimental values. Namely, in the low-spin states they have a 
rotational spectrum, but the $10^+$ and 
$12^+$ states show the diminish of the excitation energy. This change of
the spectrum is closely related to the $\alpha$+$^{40}{\rm Ca}$
clustering in the ground band and will be discussed together with its
parity doublet partner, $K^\pi$=$0^-_2$ 
band in the next subsection. 
\begin{table}
\caption{Observed (EXP)\cite{simpson1,simpson2,britz} and calculated
 (AMD) intra-band 
 $E2$ transition probabilities B($E2$;$J_i^\pi\rightarrow J_f^\pi$)
 [${\rm e}^2{\rm fm}^4$] within the ground band. For the comparison,
 $\alpha$+$^{40}{\rm Ca}$ RGM (RGM) \cite{wada}  and $\alpha$+$^{40}{\rm Ca}(I^\pi)$ OCM (OCM) \cite{ohkubo} results are also shown.} \label{tab::ground_E2} 
\begin{tabular}{l|c|c|c|c}
\hline
$J^\pi_i\rightarrow J^\pi_f$& EXP &  RGM & OCM & AMD\\\hline 
$2^+_1 \rightarrow 0^+_1$& $120\pm30$ & 107  & 166 & 142\\
$4^+_1 \rightarrow 2^+_1$& $280\pm60$ & 146 & 231 & 222 \\
$6^+_1 \rightarrow 4^+_1$& $160\pm30$  & 140 & 212 & 167 \\
$8^+_1 \rightarrow 6^+_1$& $14>$ &118 & 185 & 172\\
$10^+_1 \rightarrow 8^+_1$& $140\pm30$ &75  &138  & 99\\
$12^+_1 \rightarrow 10^+_1$& $40\pm8$  &34 & 73 & 69\\
\hline
\end{tabular}
\end{table}

\subsubsection{$K^\pi$=$3^-_1$ and $0^-_1$ bands}
The $1\hbar\omega$ excitation from the ground band produces the
$K^\pi$=$0^-_1$ and $K^\pi$=$3^-_1$ bands in the negative-parity. 
Though they have different $K$ quanta, these two bands are obtained from
the same intrinsic wave function. This induces the rather strong mixing
between the $K^\pi$=$0^-_1$ band members and the natural parity band
members of the $K^\pi$=$3^-_1$ band, and this mixing enhances the
even-odd staggering in the $K^\pi$=$3^-_1$ band. The obtained
$K^\pi$=$3^-_1$ band  corresponds to the experimental $K^\pi$=$3^-_1$
band which is 
observed from the $3^-$ state at 3.17 MeV to the $13^-$ state at 11.55
MeV. The $1\hbar\omega$ nature of this band is consistent with the
results of the shell model calculation\cite{TI44}. We have assigned the
calculated $K^\pi$=$0^-_1$ band members to the observed $1^-$ (3.76
MeV), $3^-$ (3.94 MeV) and $5^-$ (5.21 MeV or 5.31 MeV) states. In the
present calculation, member states of this band are obtained up to $9^-$
state. Though this band is not sufficiently confirmed in experiments,
the agreement between the calculation and the experiment seems to be 
satisfactory at least in the excitation energy. 

\subsubsection{$K^\pi$=$0^+_2$ and $2^+$ superdeformed bands}
In contrast to the $1\hbar\omega$ single-particle excitation in the
lowest negative-parity bands, the reconstruction of the mean-field
occurs in the positive-parity. The first excited band in the
positive-parity states is the $K^\pi$=$0^+_2$ band which mainly consists 
of the $4\hbar\omega$ structure. As is discussed in the previous
section, this 
$4\hbar\omega$ structure has a triaxially deformed intrinsic wave
function in which four nucleon are excited to the $pf$-shell from the
$^{40}{\rm Ca}$ core. Due to this many-particle excitation, the system
has a quite different shape of the mean-field and is stabilized by
the formation of the elongated superdeformed structure. After
the angular momentum projection and the GCM calculation, the excitation
energy of this band becomes quite small. It starts from the $0^+_2$
state at 1.75 MeV and reaches the $10^+$ state at 12.1 MeV in the
present calculation. The obtained excitation energies show  good
agreement with  experiments in low-spin states. As the angular
momentum increases, the observed spectrum deviates from the rotational
one, while the calculated spectrum shows the rotational character. This
may be because of the absence of the $2\hbar\omega$ 
structure in our calculation. In the shell model calculation\cite{TI44},
it is shown that the $2\hbar\omega$ structure becomes dominant as the
angular momentum increases in this band and the excitation energies of
$10^+_2$ and $12^+_2$ states are much lower than those expected from the
rotational spectrum. In our calculation, we have superposed the wave
functions on the energy curve and the $2\hbar\omega$ structure has not
appeared on the energy curve. Of course, the obtained wave function in
the present calculation is not of the pure $0\hbar\omega$ and
$4\hbar\omega$ structure, but also contains $2\hbar\omega$ structure to
some extent. However the description of the $2\hbar\omega$ structure seems
to be not sufficient in the present calculation. This may also explain
the absence of the $12^+_2$ state in our calculation. 
\begin{table}{r}{7.5cm}
\caption{Observed (EXP)\cite{simpson1,simpson2,britz} and calculated
 (AMD) intra- and inter-band  $E2$ transition probabilities
 B($E2$;$J_i^\pi\rightarrow J_f^\pi$)  [${\rm e}^2{\rm fm}^4$] within
 the $K^\pi$=$0^+_2$ and $K^\pi$=$2^+$ bands. For the comparison,   
$\alpha$+$^{40}{\rm Ca}(I^\pi)$ coupled-channel OCM (OCM) \cite{ohkubo}
 results are also  shown.} \label{tab::super_E2} 
\begin{tabular}{c|c|c|c}
\hline\hline
$K^\pi$=$0^+_2$$\rightarrow$$K^\pi$=$0^+_2$& EXP &  OCM & AMD\\\hline
$2^+_2 \rightarrow 0^+_2$& $220\pm50$   & 157 & 320 \\
$4^+_2 \rightarrow 2^+_2$& $268\pm50$   & 268 & 361  \\
\hline
$K^\pi$=$2^+$$\rightarrow$$K^\pi$=$2^+$& EXP &   OCM & AMD \\\hline
$3^+_3 \rightarrow 2^+_3$  &$<590$ & 185 & 298\\
 $4^+_3 \rightarrow 2^+_3$&$175^{+100}_{-60}$ & 148 & 220\\
 $4^+_3 \rightarrow 3^+_1$ & $<785\pm650$ & 11 & 302\\\hline
$K^\pi$=$2^+$$\rightarrow$$K^\pi$=$0^+_2$& EXP &   OCM & AMD \\\hline
 $2^+_3 \rightarrow 0^+_2$ & $43<$ & 3.04 & 24\\\hline\hline
\end{tabular}
\end{table}

Since the $4\hbar\omega$ structure has a triaxially deformed intrinsic
wave function, $K^\pi$=$2^+$ band adjoins $K^\pi$=$0^+_2$ band. The
member states of this band are obtained from the $2^+_3$ state (3.2 MeV)
up to $10^+$ (11.3 MeV) state. In the experiment, the possible existence
of the low-lying $K^\pi$=$2^+$ band which starts from the $2^+$ state at
2.89 MeV has been long known, and more recently the $8^+$, $10^+$ and $12^+$
states which possibly belong to this band have been observed together with
their $\gamma$ transitions to the $K^\pi$=$0^+_2$ band, though the
spin-parity of the $(3^+)$ state is not fixed and the $5^+$, $6^+$,
$7^+$, $9^+$ and $11^+$ states have not been observed yet. The obtained
$K^\pi$=$2^+$ band corresponds to this $K^\pi$=$2^+_1$ side
band. The calculated inter- and intra- B($E2$) values within
$K^\pi$=$0^+_2$ and $K^\pi$=$2^+_1$ bands (table \ref{tab::super_E2})
are larger than the experimental values, but show similar trend
qualitatively.

We note that these two band were also obtained up to $4^+$ states by a
semi-microscopic cluster model\cite{ohkubo} assuming the 
$\alpha$+$^{40}{\rm Ca}(0^+_2)$ and $\alpha$+$^{40}{\rm Ca}(2^+_1)$
cluster structure. This result also supports the $4\hbar\omega$
structure of these bands and means the same origin of the $K^\pi$=$0^+_2$
and $K^\pi$=$2^+$ bands, since both $^{40}{\rm Ca}(0^+_2)$ and
$^{40}{\rm Ca}(2^+_1)$ states are regarded as having the
$\alpha$+$^{36}{\rm Ar}$ cluster structure in their calculation. We also
note that their results implies that the $K^\pi$=$0^+_2$ and
$K^\pi$=$2^+$ bands are possibly regarded as having the
2$\alpha$+$^{36}{\rm Ar}$ cluster 
structure. However, in the case of our AMD result the relation between
the superdeformed state and the 2$\alpha$+$^{36}{\rm Ar}$ structure is
unclear. Indeed, in our calculation this state has a quite large
contribution from the spin-orbit force (about -27 MeV in the case of the
$0^+_2$ state) which is much larger than that expected in the case of the
2$\alpha$+$^{36}{\rm Ar}$ structure (about -14 MeV in our calculation).

\subsubsection{Hyperdeformation and $^{16}{\rm O}$+$^{28}{\rm Si}$
   molecular structure}
About 19 MeV above the ground state, we have obtained the $K^\pi$=$0^+$
band which will corresponds to the $f^8g^4$ state obtained in the
Hartree-Fock calculation\cite{inaku} with large deformation
$\beta$$\sim$0.7 and it can be classified as a
hyperdeformed state. However, as was shown in the previous section, this
state has an  $^{16}{\rm O}$+$^{28}{\rm Si}$ intrinsic structure and it
implies that it will be more suitable to understand this state as the
$^{16}{\rm O}$+$^{28}{\rm Si}$ molecular structure than to regard as the 
hyperdeformed state. Indeed this band has its counterpart in the
negative parity, $K^\pi$=$0^-$ band which also has an $^{16}{\rm
O}$+$^{28}{\rm Si}$ intrinsic wave function. 
It may be possible to regard these two bands as the parity doublet bands.  
These bands can have something to do with the possible existence of the
$^{16}{\rm O}$+$^{28}{\rm Si}$ resonant states.

\subsection{$\alpha${\rm +}$^{40}{\rm Ca}$ cluster structure}

In the previous subsection, we have shown that many kinds of the
nuclear excitation appear in the $^{44}{\rm Ti}$ and their description
by the present model shows satisfactory agreement with 
experiments. Besides these structures, the existence of the 
$\alpha$+$^{40}{\rm Ca}$ cluster structure in $^{44}{\rm Ti}$ has long
been discussed by many authors. In particular, the semi-microscopic
cluster model which uses the $\alpha$+$^{40}{\rm Ca}$ optical potential
and the $\alpha$+$^{40}{\rm Ca}$ resonating group method
calculations\cite{kihara,arima,wada} 
have shown the $\alpha$+$^{40}{\rm Ca}$ nature of the ground band and the
existence of the excited $N$=13, 14 and 15 bands of the
$\alpha$+$^{40}{\rm Ca}$ structure
negative-parity.  Here $N$ denotes the principal quantum number of the
relative motion between $\alpha$ and $^{40}{\rm Ca}$, and is defined as
$N$=$2n+L$ with $n$ and $L$ denoting the number of nodes and the angular
momentum of the $\alpha$-${\rm Ca}$ relative wave function,
respectively. $N$=12 is the lowest Pauli allowed state which corresponds
to the ground state. In this section, we discuss the $\alpha$+$^{40}{\rm
Ca}$ clustering in the obtained states on the basis of the microscopic
model, deformed-basis AMD,  which can also reproduce the excitation
mechanism other than the clustering. To clarify our argument, we show
the results of two GCM calculations which are obtained by superposing
the $0\hbar\omega$ wave functions and $\alpha$+$^{40}{\rm Ca}$ wave
functions on the energy curves in Fig. \ref{fig::positive_surface} and
\ref{fig::negative_surface} ($0\hbar\omega$+($\alpha$+$^{40}{\rm Ca}$)
GCM) and by superposing all 
of the obtained wave functions (full GCM).
The result of the $0\hbar\omega$+($\alpha$+$^{40}{\rm Ca}$) GCM
shows the existence of the non-small $\alpha$+$^{40}{\rm Ca}$ component
in the ground band and the $1\hbar\omega$, $2\hbar\omega$ and
$3\hbar\omega$ excitation of the relative motion between 
$\alpha$ and $^{40}{\rm Ca}$ in the N=13, 14 and 15 bands. The
comparison with experiments should be made by  the full GCM
calculation.  We will see that, due to the strong mixing
with the $3\hbar\omega$ structure which does not have the
$\alpha$+$^{40}{\rm Ca}$ component, N=13 and 15 bands are fragmented 
into several states in the full GCM calculation. 

\subsubsection{$\alpha$+$^{40}{\rm Ca}$ component in the ground band}
\begin{figure}
\epsfxsize =\hsize
\epsfbox{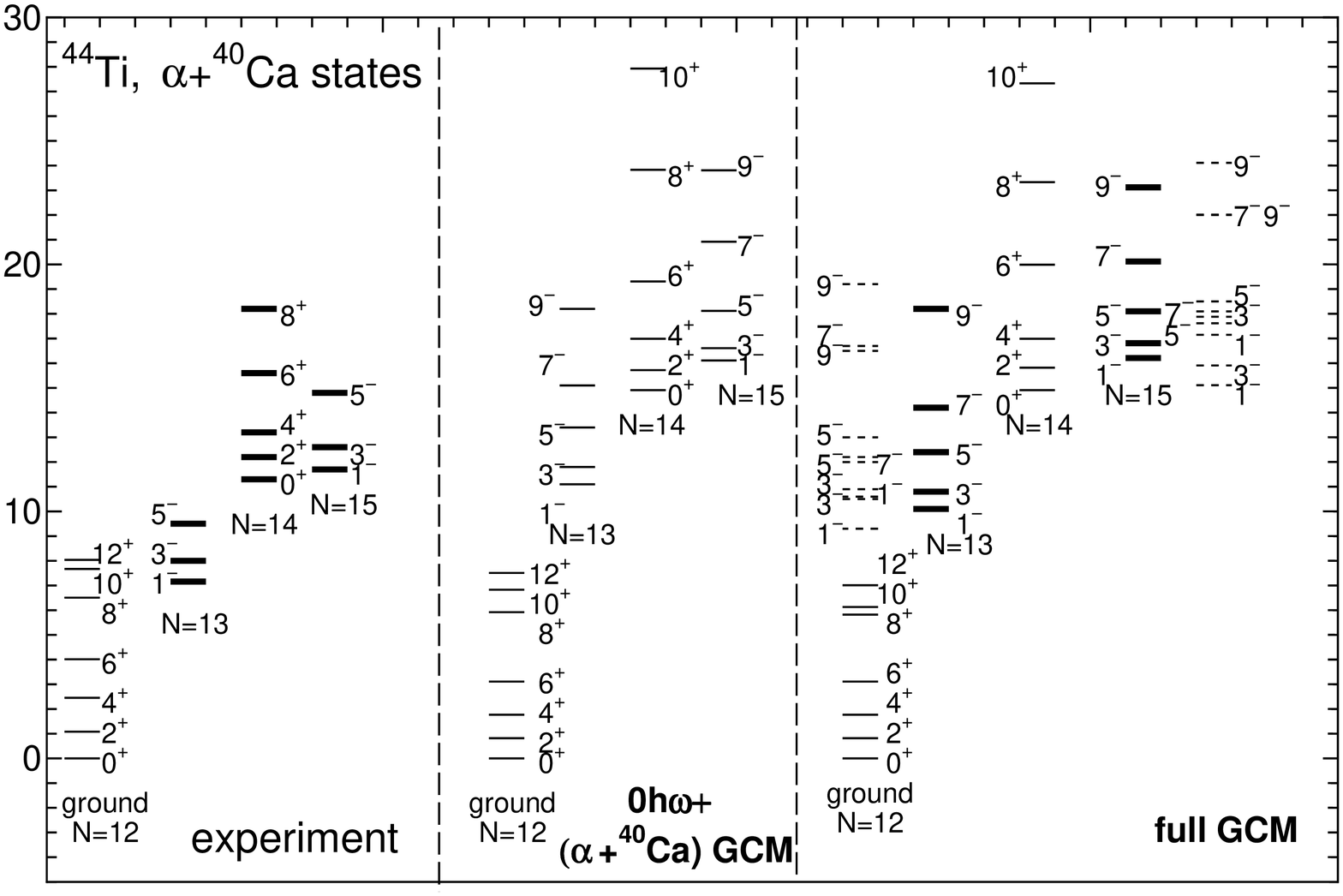}
\caption{Observed\cite{yamaya} and calculated
 ($0\hbar\omega$+($\alpha$+$^{40}{\rm Ca}$) GCM  and full GCM ) states
 of $^{44}{\rm Ti}$  which have the $\alpha$+$^{40}{\rm Ca}$ cluster
 structure.  
The bold lines indicates the energy centroid of the experimental
($N$=13,14 and 15) and theoretical (full GCM, $N$=13 and 15)
 fragmented states.  
The dotted lines indicate the energy of the fragmented states obtained
by the full GCM calculation.} 
\label{fig::level2}
\end{figure}

\begin{figure}
\epsfxsize =0.5\hsize
\epsfbox{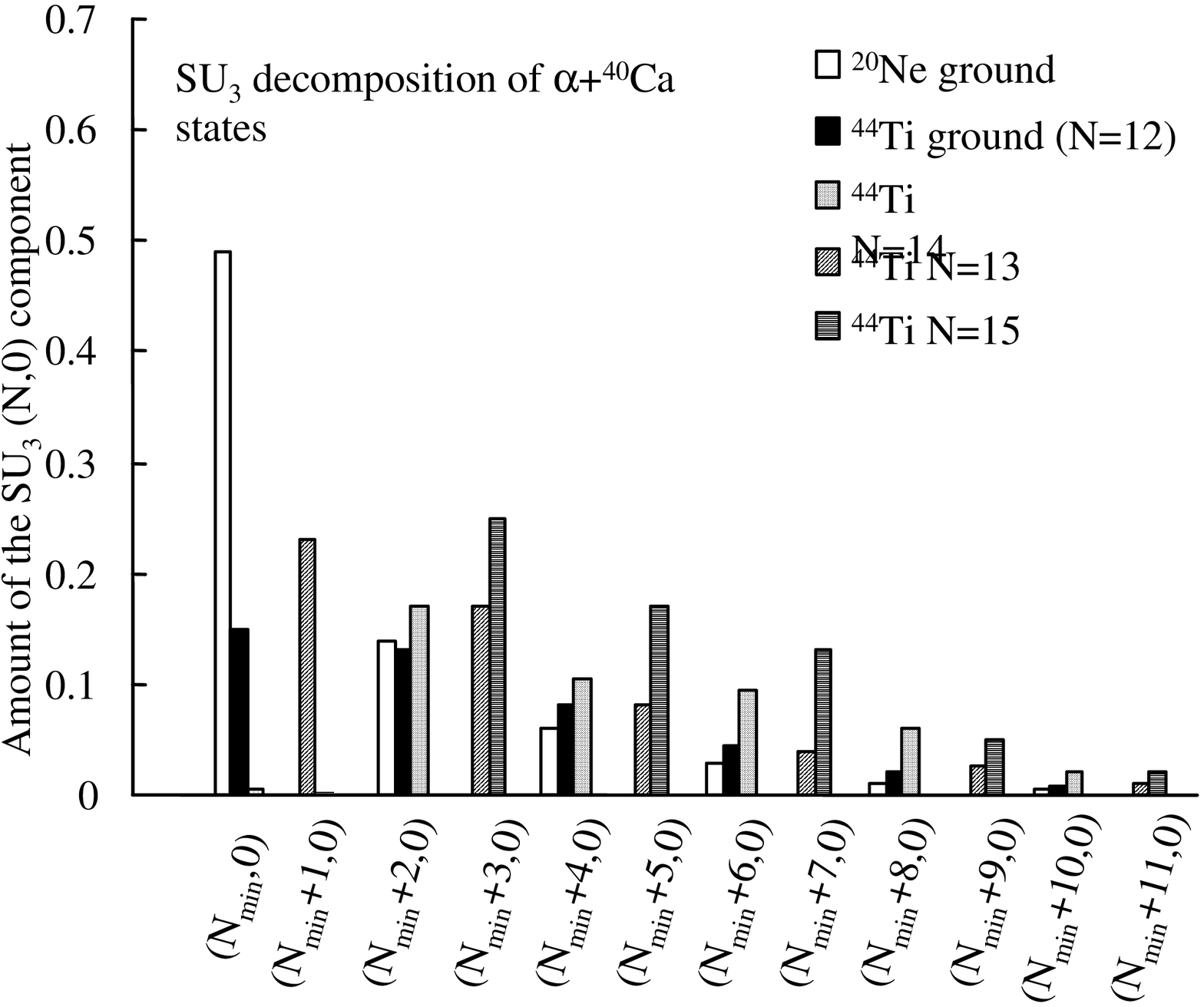}
\caption{The decomposition of the $\alpha$+$^{40}{\rm Ca}$ component
 into the $SU_3$ irreducible representations. These results are
 calculated from the $0\hbar\omega$+($\alpha$+$^{40}{\rm Ca}$) GCM wave
 functions. For comparison, the result of the ground band of $^{20}{\rm Ne}$ is also shown. $N_{min}$ indicates the lowest Pauli-allowed states. $N_{min}$=12 and $N_{min}$=8 for $^{44}$Ti and $^{20}$Ne, respectively.}
\label{fig::decomp}
\end{figure}
In the present calculation, the ground band is the only
candidate of the lowest $\alpha$+$^{40}{\rm Ca}$ band among the obtained
$K^\pi$=$0^+$ bands, since other obtained $K^\pi$=$0^+$ bands have
$2\hbar\omega$ or $4\hbar\omega$ excited structure of the $^{40}{\rm
Ca}$ core. We found that the ground band member states have non-small
amounts of the $\alpha$+$^{40}{\rm Ca}$ component except $10^+_1$ and
$12^+_1$ states.  The method of the evaluation of the
$\alpha$+$^{40}{\rm Ca}$ component in the AMD wave function is explained
in the appendix. In the case of the ground state obtained by the full
GCM calculation, it amounts to about 39\% (table \ref{tab::WJ}). However, it
is not so large as in the case of the $\alpha$+$^{16}{\rm O}$ component
in the ground state of $^{20}{\rm Ne}$\cite{kimura}, in which it amounts to
about 70\%. As is stated in the introduction, due to the increasing core
mass and the strong effect of the spin-orbit force, we cannot expect the 
prominent $\alpha$+$^{40}{\rm Ca}$ cluster structure in the ground band,
which is different from the cases of  the $sd$-shell nuclei such as
$^{16}{\rm O}$ and $^{20}{\rm  Ne}$. Indeed, the spin-orbit force has a
non-small expectation value (-9.5 MeV in the case of the ground state)
which must be zero in the case of the $\alpha$+$^{40}{\rm Ca}$
structure.  Therefore, we can conclude that the $\alpha$+$^{40}{\rm Ca}$ 
cluster structure is considerably distorted by the formation of the
deformed mean-field and the strong effect of the spin-orbit force. 

To understand the distortion of the $\alpha$+$^{40}{\rm Ca}$ cluster
structure in the ground band, we first decompose the wave function of
the ground state obtained by the full GCM calculation into the
$\alpha$+$^{40}{\rm Ca}$ cluster component and other components
orthogonal to it. Then we decompose the $\alpha$+$^{40}{\rm Ca}$ cluster
component into the $SU_3$ irreducible representations. The definitions
and  explanations of this decomposition are presented in the
appendix. Similarly to the ground state of the $^{20}{\rm Ne}$, the
ground state of $^{44}{\rm Ti}$ contains the highly excited $(N,0)$
components (figure \ref{fig::decomp}). For example,
$(N_{min}$+$8,0)$=$(20,0)$ component which corresponds to the 
$8\hbar\omega$ excitation in the spherical harmonic oscillator basis
still amounts to about 3\%. This coherent contribution from 
$(N,0)$ states is specific to the cluster correlation. However, we 
must note that the magnitude of these contribution is quite small
compared to the $^{20}{\rm Ne}$. In particular, the contribution from
the lowest Pauli-allowed state, ($(12,0)$ state in the case of the
$^{44}{\rm Ti}$ and $(8,0)$ state in the case of the $^{20}{\rm Ne}$) is
greatly diminished in the $^{44}{\rm Ti}$. This means that the
cluster structure is considerably dissolved inside the nucleus,
since smaller $N$ means the smaller inter-cluster distance. And the
residual coherent contribution from higher $(N,0)$ 
components means that the $\alpha$+$^{40}{\rm Ca}$ clustering survives
in the surface region, but its contribution is not so large as in
$^{20}{\rm Ne}$. In other
words, the binding mechanism of the ground band of $^{44}{\rm Ti}$ may
be rather dominated by the mean-field picture and the
$\alpha$+$^{40}{\rm Ca}$ cluster correlation shows up more clearly  in
the surface region  of the nucleus. It is to be noted that the ground
band still has a 
considerable amount of the $\alpha$ spectroscopic factor $S_\alpha$. Since 
the smaller $(N,0)$ components less contributes to the $S_\alpha$ due to
the smallness of the $\mu_N$ (see Eq.(\ref{eq::sfac}) in the appendix),
the dissolution of the $\alpha$+$^{40}{\rm Ca}$ component in the smaller
distance (smaller $(N,0)$ components) does not affect so much to the
$S_\alpha$. The dominance of the mean-field-like component becomes more
prominent as the angular momentum increases. In particular, the $10^+_1$ 
and $12^+_1$ states has quite small amount of $\alpha$+$^{40}{\rm Ca}$
components. Instead, in these states, the intrinsic wave functions has
an oblately deformed shape. This structure change along the yrast line
causes the irregular behavior of the excitation energies of $10^+_1$
and $12^+_1$ states. The change of the deformation to the oblate shape
near the band terminal is due to the nucleon spin alignment to the
rotation axis and was also observed in the AMD calculation of the
$^{20}{\rm Ne}$ ground band\cite{enyo}.

\begin{table}[bp]
\caption{Amount of the $\alpha$+$^{40}{\rm Ca}$ component
 $W_J$ and $\alpha$ spectroscopic factor  $S_\alpha$ of  the obtained
 states (see text and appendix). The full GCM results of N=13 and 15
 band members and the experimental value of N=13 band members show the
 sum of the fragmented states.}
\label{tab::WJ}
\begin{center}
\begin{tabular}{c|cc|cc|c||c|cc|cc} 
\hline\hline
ground & \multicolumn{2}{c}{0$\hbar\omega$+($\alpha$+$^{40}$Ca) GCM} & \multicolumn{2}{c|}{full GCM} & Exp & N=14 & \multicolumn{2}{c}{0$\hbar\omega$+($\alpha$+$^{40}$Ca) GCM} & \multicolumn{2}{c}{full GCM} \\\hline
 & $W_J$ & $S_\alpha$ &  $W_J$ & $S_\alpha$ & $S_\alpha$ &  &  $W_J$ &
 $S_\alpha$ &  $W_J$ & $S_\alpha$ \\ 
$0^+$ & 0.40  & 0.14  & 0.39  & 0.14  & 0.20  & $0^+$  & 0.48  & 0.22  & 0.46  & 0.22  \\
$2^+$ & 0.36  & 0.14  & 0.34  & 0.12  & 0.20  & $2^+$  & 0.43  & 0.23  & 0.42  & 0.23  \\
$4^+$ & 0.33  & 0.14  & 0.32  & 0.12  & 0.18  & $4^+$  & 0.38  & 0.19  & 0.38  & 0.19  \\
$6^+$ & 0.25  & 0.14  & 0.25  & 0.09  & 0.16  & $6^+$  & 0.32  & 0.17  & 0.30  & 0.17  \\
$8^+$ & 0.21  & 0.13  & 0.21  & 0.08  & 0.13  & $8^+$  & 0.23  & 0.14  & 0.21  & 0.13  \\
$10^+$ & 0.06  & 0.01  & 0.06  & 0.01  &  & $10^+$  & 0.14  & 0.08  & 0.12  & 0.08  \\
$12^+$ & 0.05  & 0.00  & 0.06  & 0.00  &  &  &  &  &  &  \\\hline
$N$=13 & \multicolumn{2}{c}{0$\hbar\omega$+($\alpha$+$^{40}$Ca) GCM} & \multicolumn{2}{c|}{full GCM} & Exp & $N$=15 & \multicolumn{2}{c}{0$\hbar\omega$+($\alpha$+$^{40}$Ca) GCM} & \multicolumn{2}{c}{full GCM} \\\hline
 &  $W_J$ & $S_\alpha$ &  $W_J$ & $S_\alpha$ & $S_\alpha$ &  &  $W_J$ & $S_\alpha$ &  $W_J$& $S_\alpha$ \\
$1^-$ & 0.52  & 0.18  & 0.56  & 0.20  & 0.25  & $1^-$  & 0.63  & 0.34  & 0.63  & 0.34  \\
$3^-$ & 0.48  & 0.16  & 0.50  & 0.18  & 0.37  & $3^-$  & 0.56  & 0.31  & 0.59  & 0.32  \\
$5^-$ & 0.41  & 0.14  & 0.43  & 0.16  & 0.30  & $5^-$  & 0.54  & 0.31  & 0.56  & 0.31  \\
$7^-$ & 0.30  & 0.10  & 0.38  & 0.12  &  & $7^-$  & 0.40  & 0.26  & 0.48  & 0.28  \\
$9^-$ & 0.28  & 0.09  & 0.32  & 0.10  &  & $9^-$  & 0.30  & 0.18  & 0.35  & 0.20
 \\\hline\hline
\end{tabular}
\end{center}
\end{table}

\subsubsection{$K^\pi$=$0^-_2$ band with $\alpha$+$^{40}{\rm Ca}$ structure}
It must be noted that the considerable distortion of the
$\alpha$+$^{40}{\rm Ca}$ cluster structure in the ground band does not
mean the less importance of the $\alpha$+$^{40}{\rm Ca}$ clustering in
this nucleus. In the previous section, we have shown that the
$1\hbar\omega$ single-particle excitation from the ground state produces
the $K^\pi$=$3^-_1$ and $K^\pi$=$0^-_1$ bands. From these state, the
single-particle excitation is regarded as the one of the basic degrees
of the freedom of the excitation in this nucleus. In the same sense, the
$\alpha$+$^{40}{\rm Ca}$ clustering is also regarded as one of the basic
degrees of freedom of the excitation, since the $K^\pi$=$0^-$ band in
which the system has a considerable amount of the $\alpha$+$^{40}{\rm
Ca}$ component is obtained in the present microscopic model calculation.
As is shown in the previous section, the wave function on the energy
curve which forms the local minimum around $\beta$=0.35 shows a
existence of the $\alpha$+$^{40}{\rm Ca}$ clustering in this
state. Indeed, when we perform the $0\hbar\omega$+($\alpha$+$^{40}{\rm
Ca}$) GCM calculation, it produces the $K^\pi$=$0^-$ band ($N$=13 band)
which starts from the $1^-$ state at 8.9 MeV and the member states of
this band have considerable amount of the $\alpha$+$^{40}{\rm Ca}$
component (table{\ref{tab::WJ}}). In the case of the $1^-$ state, it amounts
to about 60\%. This band lies in the same energy region as the
experimentally observed states. The decomposition of the relative wave  
function between $\alpha$ and $^{40}{\rm Ca}$ shows the growth of the
$\alpha$+$^{40}{\rm Ca}$ clustering in this band. Compared to the ground
band, the amount of the lowest component ((12,0) in the case of the
ground band and $(13,0)$ in the case of the $K^\pi$=$0^-_2$ band)
considerably increases together with the coherent contribution from the
higher $(N,0)$ states and they lead to the increase of the $S_\alpha$
in this band (table \ref{tab::WJ}). We can say that since the ground state
contains the non-small amount of the $\alpha$+$^{40}{\rm Ca}$
component by about 40\%, the nucleus exploits this
degrees of the freedom in the nuclear excitation. However, when we
perform the full GCM calculation, the member states of this band are
fragmented into several states because of rather strong mixing with
$3\hbar\omega$ structure which does not have $\alpha$+$^{40}{\rm Ca}$
component. In figure \ref{fig::level2}, the excitation energies of these
fragmented states and their averaged energies are plotted. The present
model space is not large enough to be directly compared with the
observed fragmented states. But, at least, we may be able to say that
one of the origin of the observed fragmentation is the mixing of the
$\alpha$+$^{40}{\rm Ca}$ structure with the $3\hbar\omega$ non
$\alpha$-clustering structure. 

\subsubsection{Excited $K^\pi$=$0^\pm$ bands with $\alpha$+$^{40}{\rm
   Ca}$ structure}
The $\alpha$+$^{40}{\rm Ca}$ clustering as a degree of freedom of the nuclear
excitation is reinforced by the existence of the excited state in which
the relative inter-cluster motion is excited and have the principal
quantum numbers $N$=14 and 15. When we perform the
$0\hbar\omega$+($\alpha$+$^{40}{\rm Ca}$) GCM, we obtain additional two
rotational bands in positive- and negative-parities. These two bands starts
from $0^+_3$ state at 15.1MeV and $1^-_3$ state at 16.9 MeV,
respectively, and they contain considerable amount of the
$\alpha$+$^{40}{\rm Ca}$ component. We confirmed that these two bands
corresponds to $N$=14 and 15 bands by investigating the relative
motion (figure \ref{fig::decomp}). Namely, $N$=14 band has a dominant
(14,0) component and $N$=15 band has a dominant (15,0) component. 
The $N$=15 band has a largest amount of the $\alpha$+$^{40}{\rm Ca}$
component among $N$=12,13,14 and 15 bands. This is because of the
largest inter-cluster  distance of this band which is due to the
$4\hbar\omega$ excitation of the relative motion.

When we perform the full GCM calculation, the $N$=15 band members also
couple with $3\hbar\omega$ structure and are fragmented. But the energy
centroids (averaged energy) are not so different as the excitation
energies of the non-fragmented states obtained by the
$0\hbar\omega$+($\alpha$+$^{40}{\rm Ca}$) GCM. The $N$=14 band has not
been fragmented in the present  calculation, though its fragmentation is
observed in the experiment. We consider that this is due to the absence
of the $2\hbar\omega$ structure in our calculation. As is already
mentioned, the $2\hbar\omega$ structure does not appear on the obtained
energy curve. And we guess that $2\hbar\omega$ structure is most likely
to be coupled with the $N$=14 band, since their principal quantum numbers
are equal.   

\section {Summary}
We have studied the low-lying level structure of $^{44}{\rm Ti}$ by
using the framework of the deformed-basis AMD with the Gogny D1S force. 
We have shown that many kinds of mutually very different nuclear
excitation appearing in this nucleus are very well reproduced unifiedly
in a single AMD framework. Namely, the single-particle
excitation in the 
$K^\pi$=$3^-_1$ and $K^\pi$=$0^-_1$ bands, the formation of the
triaxial superdeformed states in the $K^\pi$=$0^+_2$ and $K^\pi$=$2^+$
bands, and the $\alpha$+$^{40}{\rm Ca}$ clustering in some
$K^\pi$=$0^\pm$ bands. The possible existence of the $^{16}{\rm
O}$+$^{28}{\rm Si}$ molecule-like structure is also suggested in the
present calculation.  The triaxial superdeformed state suggested by 
some Hartree-Fock calculations are confirmed and it has been shown that
this superdeformed state explains the low-lying $K^\pi$=$0^+_2$ and $2^+$
bands, which suggests the interpretation of the low-lying states of
these bands as the triaxial rotor.  Among these state, we have focused
on the $\alpha$+$^{40}{\rm Ca}$ clustering in the low-energy region. We
found that the $\alpha$+$^{40}{\rm Ca}$ cluster component in the ground
band is considerably dissolved by the formation of the deformed
mean-field and the spin-orbit force. But, since the ground state still
contains non-small amount of the $\alpha$+$^{40}{\rm Ca}$ component by
about 40\%, it is possible to excite the nucleus by using the 
clustering as a degree of the freedom of nuclear excitation. This
excitation leads to the existence of the $K^\pi$=$0^\pm$ bands in which
the relative motion 
between $\alpha$ and $^{40}{\rm Ca}$ have the principal quantum number
N=13, 14 and 15. However, due to the coupling to $3\hbar\omega$
structures, $N$=13 and 15 bands members are fragmented into several
states. When we average their energies for each spin, these bands lie in
the same energy region of the observed $\alpha$+$^{40}{\rm Ca}$ bands
with N=13, 14 and 15.

\section*{Acknowledgements}
We would like to thank Dr. Y. Kanada-En'yo for valuable discussions.    
Many of the computational calculations were carried out by 
SX-5 at Research Center for Nuclear Physics, Osaka University (RCNP). 
This work was partially performed in the Research Project for Study of
Unstable Nuclei from Nuclear Cluster Aspects¡É sponsored by Institute of
Physical and Chemical Research (RIKEN). 

\appendix
\section{} 
The definition and the explanation of the $\alpha$+$^{40}{\rm Ca}$
component in the AMD wave function and its decomposition into the
$SU_3$ irreducible representations are presented. The reader is directed
to references 
\cite{ENYO_NEW,P_SHELL_CLUSTERS} for more details.  

\subsection{$\alpha${\rm +}$^{40}{\rm Ca}$ component in the AMD wave function}
The $\alpha$+$^{40}{\rm Ca}$ system is generally expressed by the
RGM-type wave function. 
\begin{eqnarray}
 \Phi_{\alpha+^{40}{\rm Ca}}^{J^\pi} &=& n_0{\cal{A}}
  \{\chi_J(r)Y_{J0}(\hat{r})\phi(\alpha)\phi(^{40}{\rm Ca})\},\label{eq::RGM0}\\    \quad n_0 &=& \sqrt{40!\cdot4!/44!}, \quad \pi = (-)^J.  
\end{eqnarray}
Here, $\cal{A}$ is the antisymmetrizer, ${\bf r}$ is the
relative coordinate between $\alpha$ and $^{40}{\rm Ca}$,
$\phi(\alpha)$ and $\phi(^{40}{\rm Ca})$ are the normalized internal wave
function of the clusters. $\chi_J(r)$ is the radial wave function of the
relative motion between $\alpha$ and $^{40}{\rm Ca}$, and is so normalized
that $\Phi_{\alpha+^{40}{\rm Ca}}^{J}$ is normalized to unity.

The normalized deformed-base AMD+GCM  wave function $\Phi_M^{J^\pi}$ of
$^{40}{\rm Ca}$ is divided into the $\alpha$+$^{40}{\rm Ca}$ component
and the residual part $\Phi_R^{J^\pi}$.
\begin{eqnarray}
 \Phi_M^{J^\pi} &=& \alpha\Phi^{J^\pi}_{\alpha+^{40}{\rm Ca}}
   + \sqrt{1-\alpha^2} \Phi_R^{J^\pi}.  \label{EQ_DECOMPOSE}
\end{eqnarray} 
$\Phi_{R}^{J^\pi}$ is also normalized and orthogonal to the
$\Phi_{\alpha+^{40}{\rm Ca}}^{J^\pi}$. 
We introduce the projection operator ${\cal P}_L$ which projects out the
$\alpha$+$^{40}{\rm Ca}$ component from the $\Phi_M^{J\pi}$,
\begin{eqnarray}
{\cal P}_L\Phi_M^{J\pi}&=&\alpha\Phi_{\alpha+^{40}{\rm Ca}}^{J\pi}\nonumber\\
                &=&\alpha n_0 {\cal A}\{\chi_J(r)Y_{J0}(\hat{r})\phi(\alpha)
		 \phi(^{40}{\rm Ca})\}.
  \label{EQ_PROJECTION} 
\end{eqnarray}
The practical formula of ${\cal P}_L$ used in this study is given
later. Using this projection operator, the squared amplitude of the
$\alpha$+$^{40}{\rm Ca}$ component (amount of the $\alpha$+$^{40}{\rm
Ca}$ component) $W_J$ of $\Phi_{M}^{J^\pi}$ is 
written as, 
\begin{eqnarray}
 W_J=|\alpha|^2 \equiv \langle \Phi^{J\pi}_M|{\cal P}_L|\Phi_M^{J\pi}\rangle.
  \label{EQ_AMOUNT}
\end{eqnarray}

\subsection{Decomposition of the $\alpha$+$^{40}{\rm Ca}$ component into
  $SU_3$ irreducible representations}

If the internal wave functions $\phi(\alpha)$ and $\phi(^{40}{\rm Ca})$
are expressed by the H.O. wave functions with oscillator parameter
$\nu$, Eq. (\ref{eq::RGM0}) and (\ref{EQ_PROJECTION}) are easily
decomposed into $SU_3$ irreducible representations. When we expand
$\chi_J(r)$ by the 
radial wave function of the H.O. $R_{NJ}(r,\gamma)$  with $N$=$2n+J$ and
the width parameter $\gamma=\frac{4\cdot 40}{44}\nu$ as   
\begin{eqnarray}
 \chi_J(r) = \sum_N e_{NJ} R_{NJ}(r,\gamma), \label{EQ_HO_EXPAND}
\end{eqnarray}
the norm of Eq. (\ref{eq::RGM0}) is written as
\begin{eqnarray}
 \langle \Phi_{\alpha+^{40}{\rm Ca}}^{J^\pi}|
\Phi_{\alpha+^{40}{\rm Ca}}^{J^\pi}\rangle = \sum_N\mu_Ne_{NJ}^2 = 1,
\end{eqnarray}
and therefore, the squared amplitude of the $\alpha$+$^{40}{\rm Ca}$ is
rewritten as 
\begin{eqnarray}
 W_J=|\alpha|^2 = |\alpha|^2\sum_N\mu_Ne_{NJ}^2.
\end{eqnarray}
and the $\alpha$ spectroscopic factor $S_\alpha$ is defined as,
\begin{eqnarray}
 S_\alpha^J\equiv|\alpha|^2\sum_N\mu_N^2e_{NJ}^2. \label{eq::sfac}
\end{eqnarray}
Here, the eigenvalue of the RGM norm kernel $\mu_N$ is defined as 
\begin{eqnarray}
 \mu_N \equiv &&\langle R_{NJ}(r)Y_{J0}({\hat r})\phi(\alpha)\phi(^{40}{\rm Ca})|   \nonumber \\
 &&{\cal A}\{R_{NJ}(r)Y_{J0}({\hat r})\phi(\alpha)\phi(^{40}{\rm Ca})\}\rangle,  \label{eq::decompose}
\end{eqnarray}
and it has a non-zero value for $N\geq12$ in the case of the
$\alpha$+$^{40}{\rm Ca}$ system.
The Eq. (\ref{eq::decompose}) means that the quantity
$|\alpha|^2\mu_Ne_{NJ}^2$ is equal to the amount of the $SU_3 (N,0)$
irreducible representations which is contained in
$\alpha\Phi_{\alpha+^{40}{\rm Ca}}^{J^\pi}$, since both $\phi(\alpha)$
and $\phi(^{40}{\rm Ca})$ 
are $SU_3$ scalar and therefore, the $\alpha$+$^{40}{\rm Ca}$ system
which has a relative motion $\chi_J(r)$=$R_{NJ}(r,\gamma)$ belongs to
$SU_3 (N,0)$.  
\subsection{Approximation of the projection operator ${\cal P}_L$}

To evaluate $W_J$ and $e_{NJ}$, we have approximated the projection
operator ${\cal P}_L$ by the set of the orthonormalized
angular-momentum-projected Brink-type wave functions. We start from the
Brink-type wave function $\varphi_B(R_i)$ in which $\alpha$ and 
$^{40}{\rm Ca}$ are located at the points $(0,0,\frac{40}{44}R_i)$ and 
$(0,0,-\frac{4}{44}R_i)$, 
respectively, 
\begin{eqnarray}
\varphi_B(R_i) &=& n_0{\cal A}\bigl\{\varphi(\alpha,\frac{40}{44}{\bf R}_i),
  \varphi(^{40}{\rm Ca},-\frac{4}{20}{\bf R}_i)\bigr\},\\
 {\bf R}_i &\equiv& (0,0,R_i).
\end{eqnarray}
For simplicity, $\varphi(\alpha)$ and $\varphi(^{40}{\rm Ca})$ are
described by the AMD wave functions which are obtained by the
variational calculation without any constraint potential and with the
fixed spherical width parameter $\nu$=$0.150$. And we have assumed that
these wave functions approximate well the $SU_3$ limits.
By separating the center-of-mass wave function $\omega({\bf X}_G)$, the
internal wave function 
$\varphi_C( R_i)$ is written as 
\begin{eqnarray}
\varphi_B(R_i)&=& \omega({\bf X}_G)\varphi_C(R_i),\\
\varphi_C(R_i)&=&n_0 {\cal A}\{\Gamma({\bf r},{\bf R}_i,\gamma)\phi(\alpha)  \phi(^{40}{\rm Ca})\},\label{EQ_BRINK_AJP}\\
\Gamma({\bf r},{\bf R}_i,\gamma)&=&\biggl(\frac{2\gamma}{\pi} \biggr)^{3/4}
e^{-\gamma({\bf r}-{\bf R}_i)^2} ,\label{EQ_BRINK_GAMMA}\\
\omega({\bf X}_G)&=&\biggl(\frac{44\cdot 2\nu}{\pi} \biggr)^{3/4}
\exp(-44\nu{\bf X}_G^2), \\
{\bf X}_G &=& \frac{1}{44}\sum_{i=1}^{44}{\bf r}_i,\quad
\gamma=\frac{40\cdot4}{44}\nu,
\end{eqnarray} 
where $\Gamma({\bf r},{\bf R_i},\gamma)$ is the relative wave function
between $\alpha$ and $^{40}{\rm Ca}$ of the Brink-type wave function.
The angular-momentum-projected Brink-type wave function $\phi^J_C(R_i)$
is obtained from the $\phi_C(R_i)$,  
\begin{eqnarray}
 \varphi_C^J(R_i) = q_i^J P^J_{00}\varphi_C(R_i).\label{EQ_APROJ_BRINK}
\end{eqnarray}
$P^J_{00}$ is the angular momentum projector and $q_i^J$ is the
normalization factor.
The orthonormalized set of the angular-momentum-projected Brink-type wave
functions $\tilde{\varphi}_\alpha^J$ is described by the linear
combination of the  $\varphi_C^J(R_i)$, 
\begin{eqnarray}
 \tilde{\varphi}_\alpha^J &=& \frac{1}{\sqrt{\rho_\alpha}}
  \sum_{i}c_{i\alpha} \varphi_C^J(R_i),
\end{eqnarray}
and the coefficients $\rho_\alpha$ and $c_{i\alpha}$ are the eigenvalues
and eigenvectors of the overlap matrix $B_{ij}=\langle \varphi_C^J(R_i)|
\varphi_C^J(R_j)\rangle$,
\begin{eqnarray}
 \sum_{j}B_{ij}c_{j\alpha} = \rho_\alpha c_{i\alpha}.
\end{eqnarray}
If the number of the basis wave functions
$\widetilde{\varphi}^{J}_\alpha$ is taken enough, the projection
operator ${\cal P}_L$ is approximated by the set of 
$\widetilde{\varphi}^J_\alpha$,
\begin{eqnarray}
 {\cal P}_L &\simeq&
  \sum_{\alpha}|\widetilde{\varphi}_\alpha^J\rangle\langle\widetilde{\varphi}_\alpha^J|,\nonumber\\ 
  &=& \sum_{ij}B^{-1}_{ji}|\varphi_C^J(R_i)\rangle\langle\varphi_C^J(R_j)|.
   \label{EQ_PROJECTOR}
\end{eqnarray}
In the present calculation, we have employed 19 Brink-wave functions in
which $R_i$ is taken as $R_i=1.0, 1.5, 2.0,..., 10.0$ fm.

Using Eq. (\ref{EQ_PROJECTOR}), Eq. (\ref{EQ_PROJECTION}) and
Eq. (\ref{EQ_AMOUNT}) are rewritten as  
\begin{eqnarray}
{\cal P}_L\Phi_M^{J\pi} &=&\alpha n_0{\cal A}\{\chi_J(r)Y_{J0}(\hat{r})\phi(\alpha)\phi(^{40}{\rm Ca})\},\nonumber\\
 &\simeq&\sum_{ij}B_{ji}^{-1}\langle \varphi_C^J({R}_j)|\Phi_M^{J^\pi}\rangle 
   |\varphi_C^J(R_i)\rangle, \label{EQ_PROJECTION2}\\
|\alpha|^2 &\simeq& \sum_{ij}\langle \varphi_C^J({R}_j)|\Phi_M^{J^\pi}\rangle
 B^{-1}_{ji} \nonumber\\
&&\times\langle \Phi_M^{J^\pi}|\varphi_C^J({R}_i)\rangle.
\end{eqnarray}

The expansion cofficient $e_N$ of $\chi_J(r)$  is
obtained from Eq. (\ref{EQ_PROJECTION2}). We expand wave function of the
relative motion $\gamma({\bf r}, {\bf R}_i, \gamma)$ of Eq. (\ref{EQ_GAM}),
\begin{eqnarray}
 \Gamma({\bf r},{\bf R_i},\gamma) = e^{-\frac{\gamma}{2}R_i^2}
  \sum_{NJm}\biggl\{\frac{(\gamma R_i^2)^\frac{N}{2}}{\sqrt{N!}}\label{EQ_GAM}\nonumber\\ 
\times\sqrt{\frac{4\pi}{2J+1}} 
  A^N_J Y_{Jm}({\hat R}_i)R_{NJ}(r,\gamma)Y_{Jm}(\hat r)\biggr\},\\
 A^N_J \equiv (-)^{(N-J)/2}\sqrt{\frac{(2J+1)\cdot N!}{(N-J)!!
  \cdot (N+J+1)!!}}.
\end{eqnarray}
And using Eq. (\ref{EQ_BRINK_AJP}) and (\ref{EQ_GAM}),
Eq. (\ref{EQ_APROJ_BRINK}) is rewritten as 
\begin{eqnarray}
 \varphi_C^J(R_i) &=& n_0{\cal A}\{\chi_J^{(i)}(r)Y_{J0}(\hat r)\phi(\alpha)
  \phi(^{40}{\rm Ca})\},\label{EQ_APROJ_BRINK2}\\
  \chi^{(i)}_J(r) &=& q_i^J e^{-\frac{\gamma}{2}R_i^2}\sum_{N}\frac{(\gamma R_i^2)^\frac{N}{2}}{\sqrt{N!}}A_J^NR_{NJ}(r).\label{EQ_BRINK_HO_EXPAND}
\end{eqnarray}
By inserting Eq. (\ref{EQ_APROJ_BRINK2}) into
Eq. (\ref{EQ_PROJECTION2}), we obtain $\chi_J(r)$ as a superposition of $\chi^{(i)}_J(r)$,
\begin{eqnarray}
\chi_J(r) \simeq \frac{1}{\alpha}\sum_{ij}\langle\varphi_C^J({R}_j)|\Phi_M^{J^\pi}\rangle B^{-1}_{ji}
  \chi^{(i)}_J(r).\label{EQ_ANS1}
\end{eqnarray}
Inserting  (\ref{EQ_BRINK_HO_EXPAND}) into Eq. (\ref{EQ_ANS1}), $e_N$
is given as 
\begin{eqnarray}
  e_N &=& \frac{1}{\alpha}A^N_J\sum_{ij}\biggl\{\langle\varphi_C^{J}({R}_j)
  |\Phi_M^{J^\pi}\rangle B^{-1}_{ji}\nonumber\\
 &&\times q_i^J e^{-\frac{\gamma}{2}{ R}^2_i}
  \frac{(\gamma {R}^2_i)^{\frac{N}{2}}}{\sqrt{N!}}\biggr\}.
\end{eqnarray}

\end{document}